\documentclass[11pt]{article}
\usepackage[text={16.5cm,23cm},centering]{geometry}

\usepackage[utf8]{inputenc}
\usepackage{tgtermes,tgheros,tgcursor}
\usepackage[varg]{newtxmath}
\usepackage{hyperref}
\hypersetup{%
 colorlinks=true,%
 linkcolor=blue,
 citecolor=blue,
}
\usepackage{xspace}
\usepackage{bm,bbm}
\usepackage{booktabs,amsmath}
\usepackage{accents}
\usepackage{multirow}
\usepackage{graphicx}
\usepackage{braket}
\usepackage{mathtools}
\usepackage{comment}
\usepackage{autobreak}
\usepackage{subcaption}
\usepackage{enumerate}
\usepackage{blkarray}
\usepackage{longtable}
\usepackage{slashed}

\usepackage{tikz}
\usetikzlibrary{patterns}
\usetikzlibrary{arrows,shapes,positioning}
\usetikzlibrary{decorations.markings}
\usetikzlibrary{decorations.pathmorphing}
\usetikzlibrary{calc}

\DeclareMathOperator{\Tr}{\mathrm{Tr}}

\def\CA{{\cal A}}

\def\d{{\rm d}}
\def\i{{\rm i}}
\def\e{{\rm e}}
\def\BZ{{\mathbb{Z}}}
\def\BR{{\mathbb{R}}}

\newcommand{\del}{\partial}

\def\KO{\mathrm{KO}}

\def\KR{\mathrm{KR}}

\def\Spin{\mathrm{Spin}}
\def\Pinp{\mathrm{Pin}^{+}}
\def\Pinm{\mathrm{Pin}^{-}}
\def\Pinpm{\mathrm{Pin}^{\pm}}
\def\DPin{\mathrm{DPin}}
\def\SU{\mathrm{SU}}
\def\SO{\mathrm{SO}}

\def\Itilde{{\text{DP}}}
\def\slope{\alpha'}
\newcommand{\Dp}[1]{\mathrm{D}{#1}}
\newcommand{\Op}[1]{\mathrm{O}{#1}}
\newcommand{\Bp}[1]{\mathrm{B}{#1}}
\newcommand{\C}{\mathrm{C}}
\newcommand{\Dnorm}[2]{\mathcal{N}_{\mathrm{D}#1,#2}}
\newcommand{\Cnorm}{\mathcal{N}_{\mathrm{C}}}

\newcommand{\antipsi}{\tilde{\psi}}
\newcommand{\antialpha}{\tilde{\alpha}}

\newcommand{\NS}{\mathrm{NS}}
\newcommand{\R}{\mathrm{R}}
\newcommand{\N}{\mathrm{N}}
\newcommand{\D}{\mathrm{D}}
\def\L{{\rm L}}
\def\time{\mathsf{T}}
\def\reflection{\mathsf{R}}
\def\fermion{\mathsf{f}}
\newcommand{\Vtach}{V_{\mathrm{tach}}}
\newcommand{\Vgauge}{V_{\mathrm{gauge}}}
\newcommand{\Vtr}{V_{\mathrm{tr}}}

\newcommand{\UI}{\mbox{$\mathrm{U}(1)$}\xspace}
\newcommand{\UIm}{\mbox{$\mathrm{U}(1)^{(m)}$}\xspace}
\newcommand{\UIw}{\mbox{$\mathrm{U}(1)^{(w)}$}\xspace}
\newcommand{\Zz}{\mbox{$\mathbb{Z}_2$}\xspace}
\newcommand{\Zzm}{\mbox{$\mathbb{Z}_2^{(m)}$}\xspace}
\newcommand{\Zzw}{\mbox{$\mathbb{Z}_2^{(w)}$}\xspace}
\newcommand{\Ndim}[1]{$#1$d\xspace}

\numberwithin{equation}{section}

\begin{document}
\begin{center}
  \begin{flushright}
    OU-HET 1252
  \end{flushright}
  \vspace{8ex}
  {\Large \bfseries \boldmath Anomalies and D-branes in the Dabholkar-Park background}\\
  \vspace{4ex}
  {\Large Hiroki Wada and Satoshi Yamaguchi}\\
  \vspace{2ex}
  {\itshape Department of Physics, Graduate School of Science, 
  \\
  Osaka University, Toyonaka, Osaka 560-0043, Japan}\\
  \vspace{1ex}
  \begin{abstract}
		We consider D-branes in the Dabholkar-Park ($\Itilde$) background, a \Ndim{9} orientifold theory obtained by gauging symmetry in the type~IIB string theory compactified on a circle.
		Using anomalies in the world-sheet theory, we provide physical insights into the classification of stable D-branes by relative KR-theory.
		The nature, such as stability, of D-branes wrapping along the compactified circle can be extracted from information about \Ndim{1} Majorana fermions on the boundary of the world-sheet.
		These Majorana fermions need to be introduced to consistently perform the GSO projection and the orientifold.
		We also construct D-brane states in the $\Itilde$ background.
		The spectrum of D-branes characterized by the relative KR-theory is correctly reproduced from the D-brane states.
	\end{abstract}
\end{center}

\section{Introduction and summary}
D-branes are important objects to obtain insights about non-perturbative aspects of string theory~\cite{Polchinski:1995mt,Polchinski:1998rr,Johnson:2023onr}.
Typically, D-branes play an essential role to understand various string dualities~\cite{Witten:1995ex,Polchinski:1995df,Horava:1995qa,Horava:1996ma}.
In particular, it is helpful to clarify the behavior of stable non-BPS D-branes under a duality in order to test the duality beyond the constraints of supersymmetry.
For this purpose, it is crucial to identify the spectrum of stable D-branes in a given string theory.
The analysis of the bound states of a brane-antibrane system with tachyon condensation and the construction of stable non-BPS states~\cite{Sen:1998rg,Sen:1998ii,Sen:1998sm,Sen:1998tt,Sen:1998ki,Sen:1998ex} motivate one to classify the charge of stable D-branes by appropriate K-theory groups~\cite{Witten:1998cd,Horava:1998jy}.
It is well-known that the KO-theory is relevant for stable D-branes in type~I theory.

Recently, it has been revealed that several properties of D-branes can be extracted from the point of view of an anomaly on the world-sheet theory\cite{Kaidi:2019pzj,Kaidi:2019tyf,Witten:2023snr}.
For instance, the GSO projection and the orientifold is needed to formulate the type~I string theory in \Ndim{10}.
These operations correspond to the gauging of the world-sheet fermion number~\cite{Seiberg:1986by,Alvarez-Gaume:1986ghj} and the world-sheet parity symmetry, respectively.
On a closed surface, we can consistently gauge these symmetries because of the absence of anomalies.
In order to introduce a D-brane, we need to formulate the world-sheet theory on a surface with boundaries.
Even if there is no anomaly on a closed surface, an anomaly may appear in the presence of boundary.
It is known that this anomaly is classified by the cyclic group $\BZ_{8}$~\cite{Fidkowski:2009dba}.
By counting zero-modes, one can identify the anomaly caused by world-sheet fermions in the presence of a boundary corresponding to each $\Dp{p}$-brane as an element of $\BZ_{8}$.
In order to formulate $\Dp{p}$-brane, we need to cancel this anomaly by adding appropriate numbers of \Ndim{1} Majorana fermion on the boundary.
The key point is that several properties of D-brane are understood by taking into account these Majorana fermions.
These results are consistent with those derived from the other approaches~\cite{Bergman:2000tm,Asakawa:2002nv,Asakawa:2002ui}.
For instance, we can study the stability of the $\Dp{p}$-brane by discussing the existence of a tachyon vertex operator depending on the number of the Majorana fermions.
Actually, one can interpret the KO-theory group classification of stable D-branes in the type~I theory in terms of anomalies and Majorana fermions~\cite{Witten:2023snr}.
In particular, the Bott periodicity of the KO-theory reflects the fact that the anomaly of the type~I theory is $\BZ_{8}$-valued.

In this paper, we explore stable D-branes in the orientifold theory in \Ndim{9}, called the Dabholkar-Park ($\Itilde$) background \cite{Dabholkar:1996pc}, based on anomalies in the world-sheet theory.
The $\Itilde$ background is defined as the orientifold of the type~IIB theory compactified on a circle.
In contrast to the type~I theory on a circle, the orientifold action involves a half-shift along the circle in the $\Itilde$ background. Due to this inclusion of the half-shift, the $\Itilde$ background is free from tadpoles without any spacetime filling D-brane.
Stable D-branes in the $\Itilde$ background are characterized by the relative KR-theory~\cite{Bergman:1999ta} shown in Table~\ref{tab:ksc}.
It is known that the T-dual of the $\Itilde$ background yields the theory on $S^{1}/\BZ_{2}$.
Here, $\BZ_{2}$ is the reflection, and $S^{1}/\BZ_{2}$ has two fixed points.
In this dual theory, one fixed point is filled by an ordinary $\Op{8}^{-}$-plane, while an $\Op{8}^{+}$-brane that has the same sign of the RR charge as the $\Dp{8}$-brane is located at the other fixed point~\cite{Witten:1997bs}.
Intuitively, D-branes near the $\Op{8}^{-}$-plane seem to be characterized by usual KO-theory, but for D-branes near the $\Op{8}^{+}$-plane the KSp-theory seems to be relevant.
In Ref.~\cite{Bergman:1999ta}, the KR-theory classification of stable D-branes in the $\Itilde$ background is qualitatively interpreted through this T-dual description.
The purpose of the present paper is to reveal the physical origin of each relative KR-theory group in the framework of the world-sheet theory of the $\Itilde$ background.

\begin{table}[thp]
	\begin{center}
	\begin{tabular}{c||cccccccccc}
		D$p$-brane & D$8$ & D$7$ & D$6$ & D$5$ & D$4$ & D$3$ & D$2$ & D$1$ & D$0$ & D$(-1)$\\\hline
		$\mathrm{KR}(S^{8-p}\times S^{1},S^{1})$ & $\BZ$ & $\BZ_{2}$ & $0$ & $\BZ$ & $\BZ$ & $\BZ_{2}$ & $0$ & $\BZ$ & $\BZ$ & $\BZ_{2}$\\
	\end{tabular}
	\end{center}
	\caption{The relative KR-theory groups that classify $\Dp{p}$-brane charges in the $\Itilde$ background.}
	\label{tab:ksc}
	\end{table}

The duality between two string theories in \Ndim{9} motivates one to study stable D-branes in the $\Itilde$ background.
The $\Itilde$ background is dual to the orbifold of the type~IIB on a circle~\cite{Dabholkar:1996pc}.
The orbifold action is $\BZ_{2}$ generated by the product of the half-shift and the left-handed spacetime fermion number.
In Ref.~\cite{Gaberdiel:2001ed}, the relation between the coupling constant $g_{\widetilde{\mathrm{I}}}$ of the $\Itilde$ background and the coupling constant $g_{\text{orbifold}}$ of the orbifold theory under this duality is specified as
\begin{align}
	g_{\text{orbifold}}=\frac{4}{g_{\widetilde{\mathrm{I}}}}.
\end{align}
This expression implies that this duality relates the weak coupling regime to the strong coupling regime and is useful to extract information beyond the perturbative analysis.
For the purpose to understand this duality, stable D-branes\footnote{The K-theory group for stable D-branes in the orbifold theory is calculated in Ref.~\cite{Gaberdiel:2001ed}. The corresponding D-brane states are constructed in Ref.~\cite{Gutperle:2000bf}} play an important role~\cite{Gaberdiel:2001ed}.
For literatures related to such \Ndim{9} string theories, see~\cite{Aharony:2007du,Vegh:2008jn,Cordova:2018dbb,Kim:2019ths,Montero:2020icj,Bena:2020qpa,Hamada:2021bbz,Bedroya:2021fbu,ParraDeFreitas:2022wnz,Montero:2022vva,Dierigl:2022reg,Etheredge:2023odp,Bossard:2024mls}.

In the $\Itilde$ background, there is massless $1$-form and $2$-form that are derived from the RR $2$-form in \Ndim{10}.
Then, $\Dp{0}$-brane and $\Dp{1}$-brane in Table~\ref{tab:ksc} are naturally understood as the $\Dp{1}$-brane wrapped or unwrapped along the circle that are electrically charged to the RR $2$-form in \Ndim{10}.
Similarly, we can interpret $\Dp{4}$-brane and $\Dp{5}$-brane in Table~\ref{tab:ksc} as the wrapped or unwrapped $\Dp{5}$-brane that are magnetically charged to the RR $2$-form in \Ndim{10}.
As discussed in Ref.~\cite{Bergman:1999ta}, the stability of non-BPS $\Dp{3}$-brane and $\Dp{7}$-brane depends on the radius $R$ of the circle.
For $R<\sqrt{2\slope}$, the $\Dp{4}$-brane and $\Dp{8}$ wrapping along the circle in \Ndim{10} are stable, while the unwrapped $\Dp{3}$-brane and $\Dp{7}$-brane become unstable.
On the other hand, for $\R>\sqrt{2\slope}$, the situation is completely reversed.
We will explain this behavior of wrapped D-branes in terms of the anomaly on the world-sheet theory.
The difference from the type~I theory on the circle comes from the contribution to the anomaly from a bosonic field in the world-sheet.
In particular, we can gain a physical understanding of the periodicity of the relative KR-theory group.
Furthermore, we explicitly construct the D-brane states and check that the constructed D-branes completely match the classification of the relative KR-theory.

The rest of this paper is organized as follows.
Section~\ref{sec:basic} is devoted to a review of the known fact about the $\Itilde$ background.
We show the spectrum of the closed string and the absence of tadpole.
The relevant K-theory for the $\Itilde$ background is also explained.
In section~\ref{sec:anomaly}, we study anomalies on the world-sheet theory.
Section~\ref{sub:typeI} is the review about analysis of the type~I theory developed in Ref.~\cite{Witten:2023snr}.
The contribution from the bosonic part is argued in section~\ref{sub:mixed_anomaly}.
Then, we discuss the stability of wrapped D-branes in section~\ref{sub:wrapped}.
After preparing several notations in section~\ref{sub:boundary_state}, we construct D-brane states in section~\ref{sub:state_unwrapped} and~\ref{sub:state_wrapped}.

\section{Spectrum of the closed string and K theory classification of D-brane}\label{sec:basic}
\subsection{Spectrum of closed string}
The $\Itilde$ background is defined as follows.
We begin with the type IIB string theory compactified on a circle $S^{1}$.
Let $X^{9}$ be the coordinate along the $S^{1}$ direction.
We denote the radius of the $S^{1}$ by $R$.
Since the world-sheet parity $\Omega$ and the half-shift along the $S^{1}$,
\begin{align}
	S:X^{9}\mapsto X^{9}+\pi R,
\end{align}
are symmetries of the type IIB theory, we can take an orientifold with $S\Omega$, and refer to this theory as the $\Itilde$ background.

In order to describe the closed string spectrum of the $\Itilde$ background, let us denote a general physical state in the closed string sector of the type IIB theory on the $S^{1}$ by $\ket{\Psi,\frac{m}{R},\frac{nR}{\slope}}$.
Here, $m$ and $n$ represent the Kaluza-Klein (KK) momentum and the winding number, respectively, while $\Psi$ represents all the other labels of the states.
Note that the half-shift $S$ acts on the state with KK momentum $m$ as $(-1)^{m}$ and acts trivially on $n$ and $\Psi$. On the other hand, the world-sheet parity $\Omega$ flips the sign of the winding number $n$, and also it may act non-trivially on $\Psi$, while it trivially acts on the KK momentum $m$.  
This implies that two types of states remain after the gauging procedure.
First ones are the states with even parity and even KK momentum written as
\begin{align}
	\Ket{\Psi,\frac{2m}{R},\frac{nR}{\slope}}+\Ket{\Omega\Psi,\frac{2m}{R},-\frac{nR}{\slope}}.
\end{align}
The second type consists of the odd parity states with odd KK momentum written as
\begin{align}
	\Ket{\Psi,\frac{2m+1}{R},\frac{nR}{\slope}}-\Ket{\Omega\Psi,\frac{2m+1}{R},-\frac{nR}{\slope}}.
\end{align}
Since the orbifold projection behaves as $\frac{1}{2}(1+\Omega)$ for even momentum states, the spectrum of these states in the $\Itilde$ background is the same as that of the type~I theory compactified on a circle.
In particular, the massless states of these theories are identical.
On the other hand, the spectrum of odd momentum states is different.
For instance, in contrast to the type~I theory, the RR $2$-form with odd KK momentum is eliminated in the $\Itilde$ background.

For the $\Itilde$ background, the contribution from the RR sector to the vacuum amplitude of the Klein bottle is given by
\begin{align}\begin{aligned}
	\CA_{K_{2},\R\R}
	&=\int_{0}^{\infty}\,\frac{\d t}{2t}\,\Tr_{\NS\NS}\left(\frac{(-1)^{\fermion_{\L}}}{2}\cdot\frac{1}{2}+\frac{1}{2}\cdot\frac{(-1)^{\fermion_{\R}}}{2}\right)\cdot\frac{S\Omega}{2}\e^{-2\pi t H_{\mathrm{c}}}\\
	&=-2^{3}V_{9}(4\pi^{2}\slope)^{-\frac{9}{2}}\left(\frac{R^{2}}{\slope}\right)^{\frac{1}{2}}\int_{0}^{\infty}\,\d s\,\vartheta_{10}\left(\frac{4\i R^{2}s}{\slope}\right)\eta(2\i s)^{-8}f_{2}(2\i s)^{8}\\
	&=-2^{3}V_{9}(4\pi^{2}\slope)^{-\frac{9}{2}}\left(\frac{R^{2}}{\slope}\right)^{\frac{1}{2}}\int_{0}^{\infty}\,\d s\,2^{5}\exp\left(-\frac{\pi R^{2}s}{\slope}\right)+\cdots.
\end{aligned}\end{align}
The theta function and $f_{2}$ are defined in Eq.~\eqref{eq:theta} and Eq.~\eqref{eq:f_func}.
This expression shows that there is no tadpole and that no spacetime filling D-brane is included to cancel the tadpole.

\subsection{Classification of D-brane charge}
In this subsection, we specify the appropriate K-theory to classify the stable D-branes in the $\Itilde$ background and identify these K-theory groups explicitly.

It is known that the stable D-branes in the type IIB theory on spacetime $X$ are classified by complex K-theory~\cite{Witten:1998cd}.
An element of the group $\mathrm{K}(X)$ is an equivalence class of pairs of bundles $(E,F)$ over $X$.
This representative corresponds to a brane-antibrane system on which $E$ and $F$ arise as gauge bundles.
A pair $(E,F)$ is defined to be equivalent to a pair $(E', F')$ if there exists bundles $H$ and $H'$ over $X$ such that $(E\oplus H, F\oplus H)$ is isomorphic to $(E'\oplus H',F'\oplus H')$.
The definition of this equivalence relation reflects the fact that a collection of branes and antibranes with the same bundle $H$ is created or annihilated by the tachyon condensation, provided that we are concerned with the conserved charge.
Although $X$ may be non-compact, the D-brane charge is classified by $\mathrm{K}$-theory with compact support, since we are interested in a finite energy configuration.
In particular, the $\Dp{p}$-brane charge is regarded as taking values in $\mathrm{K}(S^{9-p},\{\text{pt}\})$.

We next discuss D-branes in the type IIB compactified on $d$-dimensional manifold $Y$.
In order to classify $\Dp{p}$-branes on $\BR^{10-d}\times Y$ in $(10-d)$-dimensional sense, it is relevant to consider a pair of bundles over $\BR^{9-d-p}\times Y$.
As mentioned above, our target is finite energy configuration.
Thus, we add a point at infinity to obtain $S^{9-d-p}\times Y=(\BR^{9-d-p}\cup\{\infty\})\times Y$, and impose the condition that the pair of bundles is trivial at infinity $\{\infty\}\times Y$~\cite{Bergman:1999ta}.
The equivalence classes of such pairs of bundles constitute a relative K-theory group denoted by $\mathrm{K}(S^{9-d-p}\times Y,Y)$.

It is helpful to specify the actions of the world-sheet parity $\Omega$ and the half-shift $S$ on a bundle over $\BR^{9}\times S^{1}$ to generalize the above discussion to the $\Itilde$ background.
Since the parity transformation exchanges the Chan-Paton labels at the two ends of the open string, $\Omega$ acts on the bundle $E$ on the brane as complex conjugation, $E\mapsto\overline{E}$.
On the other hand, the half-shift $S$ induces the pull-back map~$S^{\ast}$.
Then, in the $\Itilde$ background, the bundles that satisfy $S^{\ast}(E) \cong \overline{E}$ are relevant.
We consider the pair of such bundles $(E,F)$ is equivalent to $(E\oplus H, F \oplus H)$ for a bundle $H$ that also satisfy $S^{\ast} (H) \cong \overline{H}$. The group composed of the classes defined by this equivalence relation is called KR-group.
The stable branes in a theory constructed from an orientifold accompanied by involution are captured by KR-group~\cite{Witten:1998cd}.
For the $\Itilde$ background, we consider relative KR-group\footnote{As mentioned in Ref.~\cite{Bergman:1999ta}, it is known in the mathematics literature that the KR-theory~$\mathrm{KR}(X\times S^{1})$ for the $\Itilde$ background in the spacetime $X\times S^{1}$ is isomorphic to the self-conjugate K-theory $\mathrm{KSC}(X)$.}.
More concretely, the $\Dp{p}$-branes in nine-dimensional term are classified by the group $\mathrm{KR}(S^{8-p}\times S^{1},S^{1})$~\cite{Bergman:1999ta}, listed in Table~\ref{tab:ksc}.
Note that these groups have the 4-fold periodicity
\begin{align}
	\KR(S^{8-(p+4)}\times S^{1},S^{1})\cong\mathrm{KR}(S^{8-p}\times S^{1},S^{1}),
\end{align}
in contrast to the 8-fold periodicity of the $\KO$-groups that capture the nature of the D-branes in the type~I theory.

\section{Anomaly of world-sheet theory}\label{sec:anomaly}
\subsection{Anomalies from world-sheet fermions}\label{sub:typeI}
In order to understand the origin of the spectrum of the stable D-branes shown in Table.~\ref{tab:ksc} and construct the boundary states for these branes, we carefully analyze the world-sheet theory of the $\Itilde$ background.
In the world-sheet point of view, the closed string sector of the $\Itilde$ background is obtained by starting from the world-sheet theory on the closed Riemann surface for the type~IIB with $(1,1)$~supersymmetry.
Then, we take the GSO projection and gauge the symmetry generated by the operation combined the world-sheet parity $\Omega$ and the half-shift $S$ along the $X^{9}$ direction.
Inserting D-branes corresponds to introducing the open string sector.
In other words, we formulate the world-sheet theory on the surface with the boundary.

Even if the theory is well-defined on the closed manifold, the theory may become inconsistent in the presence of the boundaries.
In our current setup, this inconsistency typically arises from the zero-modes of the world-sheet fermions.
If these zero-modes have anomalies of the fermion number symmetry and $S\Omega$, the GSO projection and the gauging of $S\Omega$ cannot be performed consistently.
In such a case, we need to introduce the appropriate degrees of freedom to cancel these anomalies.

Since the half-shift $S$ acts trivially on the world-sheet fermions, the anomalies from the world-sheet fermions in the $\Itilde$ background is the same as that of the type~I theory.
Hence, we first review the construction of the D-branes in the type~I theory based on the anomalies on the world-sheet, developed in Ref.~\cite{Witten:2023snr}.

To this end, we first specify the structure group for the fermion fields on the world-sheet.
For the type~IIB theory, which is the starting point to construct the type~I theory, the structure group is $\Spin(2)\times\BZ_{2}$.
The structure group in type~IIB theory includes the $\BZ_{2}$ factor, since we can independently choose the spin structure for the left-moving part and for right-moving part, in contrast to the type~$0$ theory.
Then, we sum over the contributions from all spin structures and the $\BZ_{2}$ bundles for each orientable world-sheet.
This procedure is nothing but the GSO projection.
By the orientifold to construct the type~I theory, unorientable surfaces are taken into account.
The structure group of these surfaces for the type~I theory is called $\DPin(2)$~\cite{Kaidi:2019tyf}.
The significant property of $\DPin(2)$ is that this group includes $\Pinp(2),\Pinm(2),$ and, $\Spin(2)\times\BZ_{2}$ as subgroups.
The groups $\Pinpm(2)$ are two different double cover of $\mathrm{O}(2)$.
In $\Pinp(2)$, the time reversal $\time$ and the spatial reflection $\reflection$ are lifted such that
\begin{align}
	\time^{2}&=(-1)^{\fermion},&\reflection^{2}&=1,
\end{align}
where $(-1)^{\fermion}$ is the world-sheet fermion number.
On the other hand, in $\Pinm(2)$, these two elements are lifted such that
\begin{align}
	\time^{2}&=1,&\reflection^{2}&=(-1)^{\fermion}.
\end{align}
In order to consistently perform the GSO projection and the orientifold in the presence of the D-branes, it is necessary that the world-sheet fermion number $(-1)^{\fermion}$ and the time reversal $\time$ are correctly represented on Hilbert space for the open string, which is defined on the interval with the boundary condition corresponding the D-branes.
In the rest of this subsection, we focus on the time reversal $\time$ that satisfies $\time^{2}=1$, since the group $\DPin(2)$ includes this operation.
Our next goal is to identify the anomaly of the world-sheet fermion number $(-1)^{\fermion}$ and the time reversal $\time$ such that $\time^{2}=1$ in the presence of the boundary corresponding to the $\Dp{p}$-brane.
If there exists the anomaly, we couple the Majorana fermions on the boundary to cancel the anomaly.

In Ref.~\cite{Witten:2023snr}, it is argued that the anomaly caused by the world-sheet fermions in the presence of the $\Dp{p}$-brane is the same as that arising from $(9-p)$ one-dimensional Majorana fermions $\chi^{i}$, 
which satisfy $\time\chi^{i}\time^{-1}=-\chi^{i}$, on the boundary.\footnote{Here, we focus on the boundary at $\sigma = 0$. The boundary at $\sigma = \pi$ can be analyzed similarly, 
but the behavior of the boundary fermions under $\time$ is opposite to that at $\sigma = 0$.}
Since the anomaly associated with the world-sheet fermion number $(-1)^{\fermion}$ and time-reversal symmetry $\time$ in \Ndim{1} is classified by $\BZ_{8}$~\cite{Fidkowski:2009dba}, only $(9-p)$ modulo $8$ is relevant.
This anomaly can be compensated by adding $(9-p)$ one-dimensional Majorana fermions $\lambda^{i}$ on the boundary with the transformation $\time\lambda^{i}\time^{-1}=\lambda^{i}$.
We can also cancel this anomaly by introducing $(p-1)$ Majorana fermions $\rho^{i}$ on the boundary that is odd under the time reversal $\time$, because the system consisting of $8$ Majorana fermions in \Ndim{1} with the same behavior under $\time$ has no anomaly.
Hence, the $\Dp{p}$-brane can be formulated by coupling at most $4$ Majorana fermions.
As discussed in Ref.~\cite{Witten:2023snr}, the symplectic Chan-Paton factor can compensate the anomaly arising from 4 Majorana fermions with the same transformation property.
One prepares a Hilbert space of dimension $2$ for the Chan-Paton degrees of freedom.
On this space, the fermion number $(-1)^{\fermion}$ trivially acts, and the time reversal $\time$ is defined as
\begin{align}
	\time&=\star M,&M&=\begin{pmatrix}0&-1\\1&0\end{pmatrix},
\end{align}
where $\star$ stands for the complex conjugation.
With this definition, $\time$ satisfies $\time^{2}=-1$, and this anomaly corresponds to the generator of the $\BZ_{2}$ subgroup of $\BZ_{8}$.
For instance, the $\Dp{4}$-brane in the type~I in \Ndim{10} can be formulated either by introducing $3$ Majorana fermions $\rho_{1}$, $\rho_{2}$, and $\rho_{3}$ such that $\time\rho_{i}\time^{-1}=-\rho_{i}$ or by introducing the symplectic Chan-Paton factor and a single Majorana fermion $\lambda$ such that $\time\lambda\time^{-1}=\lambda$.

In order to elucidate the nature of the D-branes, we need to study the open-string vertex operators.
For this purpose, it is convenient to regard the orientifold for the $\Itilde$ background as the gauging of the combination of the reflection $\reflection$ and the half-shift $S$.
The stability of the D-brane can be argued, using the vertex operator for a tachyon state.
We denote the tachyon vertex operator constructed from the fields in the bulk of the world-sheet as $\Vtach$.
The Majorana fermions on the boundary, the Chan-Paton factor, and the KK momentum are not included in $\Vtach$.
The operator is odd under $(-1)^{\fermion}$, while it is even under $\reflection$.
The D-brane is unstable, if we render this operator invariant under the GSO projection and the orientifold by combining it with appropriate degrees of freedom on the boundary and the KK momentum.
In the subsequent analysis, we also discuss the vertex operators for the gauge fields and for the transverse motion of the D-brane.
We denote the former one by $\Vgauge$ and the latter one by $\Vtr$.
Again, $\Vgauge$ and $\Vtr$ do not contain any degrees of freedom on the boundary and the KK momentum.
These two operators are both bosonic.
The gauge field $\Vgauge$ is odd under $\reflection$, while $\Vtr$ is even.

\subsection{Anomalies from world-sheet bosons}\label{sub:mixed_anomaly}
In ordinary type~I theory, world-sheet bosons do not contribute the anomaly.
From this fact, the classification by KO-theory can be understood only by the anomalies from the world-sheet fermions~\cite{Witten:2023snr}.
On the other hand, the orientifold action of the $\Itilde$ background is accompanied with the half-shift $S$ along $S^{1}$.
Since half-shift $S$ non-trivially acts on the world-sheet boson $X^{9}$, not only the world-sheet fermions but also the world-sheet boson can affect the anomaly.

In an orientifold theory, we take into account unorientable world-sheets on which the first Stiefel-Whitney class is non-trivial.
By the effect of the half-shift $S$, the configuration of the compact boson $X^{9}$ is constrained so that $S$ acts on $X^{9}$ along orientation reversal cycle
Note that the half-shift generates the $\Zz$ subgroup of the $U(1)$ momentum symmetry. This \UI momentum symmetry and its \Zz subgroup are denoted by \UIm and \Zzm , respectively.  
Then, the bosonic part along the $X^{9}$ direction of the world-sheet theory in the world-sheet $\Sigma$ for the $\Itilde$ background is obtained by introducing the background gauge fields $A_{m}\in H^{1}(\Sigma,\BZ_{2})$ for \Zzm and identifying $A_{m}$ with $w_{1}(\Sigma)$, where $w_{1}(\Sigma)$ is the first Stiefel-Whitney class of $\Sigma$.

The compact boson theory also has the winding \UI symmetry, denoted by \UIw.
There is a mixed anomaly between the \UIm symmetry and the \UIw symmetry.
For our current purpose, it is enough to focus on the $\Zzm \times \Zzw$ subgroup of the $\UIm \times \UIw$ symmetry.
In terms of the invertible field theory in three dimensions, the mixed anomaly of this subgroup is given by
\begin{align}
	S=\int A_{w}\smallsmile\mathrm{Bock}(A_{m})
\end{align}
where $\mathrm{Bock}$ is the Bockstein homomorphism of the short exact sequence $\BZ_{2}\to\BZ_{4}\to\BZ_{2}$, and $A_{m},\,A_{w}$ are the $\BZ_{2}$-valued gauge fields of \Zzm and \Zzw symmetries, respectively.
The existence of this anomaly implies that the \Zzw winding symmetry becomes anomalous, since the background gauge field $A_{m}$ is coupled to $X^{9}$.
The anomalous winding transformation yields the topological term
\begin{align}\label{eq:inv_phase}
	\int_{\Sigma}\mathrm{Bock}(A_{w})=\int_{\Sigma}\mathrm{Bock}(w_{1}(\Sigma))=\int_{\Sigma}w_{1}(\Sigma)^{2},
\end{align}
on the world-sheet\footnote{The Bockstein homomorphism is $\mathrm{Sq^{1}}$, where $\mathrm{Sq^{1}}$ is the Steenrod square. For $x\in H^{1}(\Sigma,\BZ_{2})$, $\mathrm{Sq^{1}}(x)=x\smallsmile x$. See e.g.~chapter 4.~L. in Ref.~\cite{Hatcher:2002}.}.  Thus, there is a unique $\Itilde$ orientifold in contrast to ``type I'' theory, where there are two distinct possible orientifolds O$9^{\pm}$ depending on the existence of the topological term \eqref{eq:inv_phase} \cite{Kaidi:2019pzj,Kaidi:2019tyf}.
On the world-sheet with boundary, this invertible phase produces the anomaly of time reversal $\time$, that is canceled by four Majorana fermions on the boundary~\cite{Kaidi:2019tyf}.
Thus, we can shift the anomaly coefficient by four units.
This phenomenon is similar to the fact that the theta term of the \Ndim{4} massless QCD is not physical, since the theta angle is shifted by the anomalous axial \UI rotation.

In the string theory in \Ndim{10} obtained by the gauging of the pure time reversal, only the world-sheet fermions contribute the anomaly, and the anomaly is $\BZ_{8}$-valued as mentioned in the previous subsection~\ref{sub:typeI}.
As the result, the periodicity of the K-theory groups that classify the stable branes in the type~I theory are eight.
In this case, the topological phase~\eqref{eq:inv_phase} is the discrete torsion associated with the gauging of the time reversal symmetry.
In other words, the two choices of the $\BZ_{2}$-valued parameter in front of the topological term~\eqref{eq:inv_phase} yield two different string theories.
The classification of the stable D-branes is given by KO-theory in the absence of the topological term, while the classification becomes KSp-theory, in the presence of the topological term.
In both cases, the periodicity of the corresponding K-theory is eight.

On the other hand, in the $\Itilde$ background, the $\BZ_{8}$-valued anomaly can be freely shifted by four units, since the topological term~\eqref{eq:inv_phase} is not physical.
As the result, the periodicity of the corresponding K-theory reduces to four\footnote{In Ref.~\cite{Bergman:1999ta}, this nature of the periodicity is physically interpreted from another point of view. In this literature, the authors utilize the fact that the $\Itilde$ background is T-dual to the theory with a $\Op{8}^{-}$-plane and a $\Op{8}^{+}$-plane~\cite{Witten:1997bs}.}, as can be observed in Table~\ref{tab:ksc}.

\subsection{Wrapped branes}\label{sub:wrapped}
There are two candidates for the $\Dp{p}$-branes in the $\Itilde$ background in \Ndim{9}.
The first one is the $\Dp{(p+1)}$-brane in \Ndim{10} wrapped along $S^{1}$.
We refer to such a brane as the wrapped $\Dp{p}$-brane.
The second one is the unwrapped $\Dp{p}$-brane in the ten-dimensional sense.
In this subsection, we discuss each wrapped $\Dp{p}$-brane in the $\Itilde$ background based on the anomaly on the world-sheet.

\paragraph{D8-branes and D0-branes:}
Since the wrapped $\Dp{8}$-branes and wrapped $\Dp{0}$-branes in the $\Itilde$ background are $\Dp{9}$-branes and $\Dp{1}$-branes in terms of \Ndim{10}, the anomaly on the world-sheet is absent for these branes.
These branes have integer valued RR charges.

\paragraph{D7-branes and D(--1)-branes:}
In these cases, the anomaly on the world-sheet is canceled by introducing a single one-dimensional Majorana fermion $\lambda$ such that $\time\lambda\time^{-1}=+\lambda$.
On these branes, we can consider the vertex operator
\begin{align}\label{eq:tach_D7}
\lambda \Vtach\e^{\pm\i\frac{1}{R}X^{9}}.
\end{align}
This vertex operator is invariant under the world-sheet fermion number $(-1)^{\fermion}$, since both $\lambda$ and $\Vtach\e^{\pm\i\frac{1}{R}X^{9}}$ are fermionic.
As explained in Ref.~\cite{Witten:2023snr}, this vertex operator is odd under the reflection $\reflection$.
Note that the $\Itilde$ background is defined by the orientifold accompanied with the half-shift $S$, and the vertex operator~\eqref{eq:tach_D7} carries an odd KK momentum.
These facts imply that the vertex operator~\eqref{eq:tach_D7} survives under the orientifold for the $\Itilde$ background.
The mass of the physical state corresponding to this vertex operator is given by
\begin{align}
	M^{2}=\left(\frac{1}{R}\right)^{2}-\frac{1}{2\slope}.
\end{align}
For $R>\sqrt{2\slope}$, this state becomes a tachyon.
Then, the branes are rendered unstable.
On the other hand, the branes are stable for $R<\sqrt{2\slope}$, since the tachyon disappears.
Because of the anomalous fermion number symmetry of $\lambda$, we cannot distinguish between a brane and an antibrane.
This implies that the classification is given by $\BZ_{2}$.

\paragraph{D6-branes:}
To construct a wrapped $\Dp{6}$-brane, we need to add two Majorana fermions $\lambda_{1}$ and $\lambda_{2}$ such that $\time\lambda_{i}\time^{-1}=+\lambda_{i}$.
In this case, we can construct the following vertex operator without the KK momentum, that is invariant under the GSO projection and the orientifold,
\begin{align}\label{eq:D6_U1}
	\lambda_{1}\lambda_{2}\Vgauge.
\end{align}
Thus, the single wrapped $\Dp{6}$-brane supports a $\mathrm{U}(1)$ gauge field.
We can introduce a Wilson loop for this $\mathrm{U}(1)$ gauge symmetry.

Note that since the state associated with the vertex operators
\begin{align}\label{eq:tach_D6}
	\lambda_{i}\Vtach\e^{\pm\i\frac{1}{R}X^{9}}
\end{align}
with a unit KK momentum are charged under the $\mathrm{U}(1)$ symmetry generated by the operator~\eqref{eq:D6_U1}, the Wilson loop shifts the mass of these states.
Similar to the operator~\eqref{eq:tach_D7}, these vertex operators~\eqref{eq:tach_D6} are invariant under the GSO projection and the orientifold, and the corresponding states are physical.
Then, after taking appropriate linear combinations and choosing normalization of the $\mathrm{U}(1)$ group, one can obtain the state with the mass
\begin{align}
	M^{2}=\left(\frac{1}{R}-\frac{\theta}{\pi R}\right)^{2}-\frac{1}{2\slope}.
\end{align}
Here, $\theta$ is a continuous parameter of the Wilson loop.
In particular, for $\theta=\pi$, such a state becomes a tachyon independent of the radius.
Since $\theta$ is a continuous parameter, even if the tachyon seems to be absent for a specific choice of the radius, we can render the $\Dp{6}$-brane unstable by the continuous deformation in the moduli space of the open string.
As a result, there is no stable $\Dp{6}$-brane in the $\Itilde$ background.

\paragraph{D5-branes:}
On the boundary of the world-sheet corresponding to the wrapped $\Dp{5}$-brane, we prepare three Majorana fermions $\lambda_{1}$, $\lambda_{2}$, and $\lambda_{3}$, which are even under the time reversal $\time$.
By utilizing all three Majorana fermions, we can construct the vertex operator,
\begin{align}
	\lambda_{1}\lambda_{2}\lambda_{3}\Vtach,
\end{align}
without the KK momentum.
This operator corresponding to the tachyon survives as the physical state in the $\Itilde$ background, since it is even under $(-1)^{\fermion}$, $\reflection$, and $S$.
Then, we conclude that the wrapped $\Dp{5}$-brane is unstable.

\paragraph{D4-branes:}
The anomaly that appears in the presence of the wrapped $\Dp{4}$-brane can be canceled by introducing the Chan-Paton degrees of freedom on which the time reversal acts as $\time^{2}=-1$.
In other words, the symplectic Chan-Paton factor is coupled to the boundary of the world-sheet corresponding to such a wrapped $\Dp{4}$-brane.
On the other hand, we can shift the anomaly by four units, as mentioned in the previous subsection~\ref{sub:mixed_anomaly}.
After this shift, the anomaly on the world-sheet is removed.
In this description, we do not need to introduce the Chan-Paton factor and the Majorana fermions on the boundary.
For the stack of this type of branes, we couple the usual orthogonal Chan-Paton factor, on which the time reversal $\time$ acting satisfies $\time^{2}=1$.

To clarify the relation between this description and the original description with the symplectic Chan-Paton factor, we consider the stack of two wrapped $\Dp{4}$-brane with the orthogonal Chan-Paton factor.
The Hilbert space for the Chan-Paton degrees of freedom is a vector space of dimension $2$, and we represent the operators acting on this Hilbert space as $2\times 2$ matrices.
The vertex operator for the gauge field on the brane is given by
\begin{align}
	\mathsf{a} \Vgauge.
\end{align}
The $2\times 2$ matrix $\mathsf{a}$ acts on the Hilbert space of the Chan-Paton degrees of freedom.
For this operator to be massless, we set the KK momentum to be zero.
The reflection $\reflection$ maps $\mathsf{a}$ to its transpose.
Since the operator $\Vgauge$ is odd under $\reflection$, the matrix $\mathsf{a}$ needs to be antisymmetric.
To make our discussion concrete, we adopt the identity matrix $I$ and the Pauli matrices $\sigma_{i}$ as the basis of the $2\times 2$ matrices.
In terms of this basis, the vertex operator for the massless gauge field is written as $\sigma_{2}\Vgauge$.
Hence, the stack of two wrapped $\Dp{4}$-brane supports the $\SO(2)\simeq\mathrm{U}(1)$ gauge symmetry.
This observation implies that the Wilson loop for this $\mathrm{U}(1)$ gauge symmetry can be introduced.

The vertex operator which describes the transverse motion of the D-branes takes the form $\mathsf{a}\Vtr$.
Here, we again focus on the vertex operator without the KK momentum.
The operator $\Vtr$ is even under the reflection $\reflection$, in contrast to $\Vgauge$.
Then, $\mathsf{a}\Vtr$ survives the orientifold projection, when $\mathsf{a}$ is symmetric.
That is for each normal direction, there are three independent ways to move the branes corresponding to $\mathsf{a}=I,\,\sigma_{1},\,\sigma_{3}$.
The vertex operator $I\Vtr$ describes the overall transverse motion of the stack of the D-branes.
On the other hand, $\sigma_{1}\Vtr$ and $\sigma_{3}\Vtr$ correspond to the transverse motions in which the stack of the D-branes separates into two components.
This means that the stack of two wrapped $\Dp{4}$-branes with the orthogonal Chan-Paton factor can be displaced from each other in the transverse directions.

We next turn on the Wilson loop for the $\mathrm{U}(1)$ gauge symmetry.
Since the generator of the $\mathrm{U}(1)$ gauge symmetry is given by $\frac{1}{2}\sigma_{2}$, the vertex operators with $I$ or $\sigma_{2}$ do not couple to the $\mathrm{U}(1)$ gauge field, while the vertex operators proportional to $\sigma_{3}\pm\i\sigma_{1}$ have the $\mathrm{U}(1)$ charge.
Even after the insertion of the Wilson loop, the state described by the vertex operator $I\Vtr$ remains massless.
The existence of this massless state implies that the overall transverse motion is allowed for any value of the $\mathrm{U}(1)$ holonomy.
On the other hand, the states corresponding to the vertex operator involving $\sigma_{3}\pm\i\sigma_{1}$ acquire the non-zero mass.
By choosing the appropriate normalization for the $\mathrm{U}(1)$ symmetry, the mass for the operators consisting of $\Vtr$ and $\sigma_{3}\pm\i\sigma_{1}$ and carrying the KK momentum $m$ is given by
\begin{align}\label{eq:mass_wboson}
	M^{2}=\left(\frac{m}{R}\mp\frac{\theta}{\pi R}\right)^{2},
\end{align}
where $\theta$ parametrizes the $\mathrm{U}(1)$ holonomy.
Now, we focus on the case of $\theta=\pi$.
These states are invariant under the orientifold for even $m$, but such states become massive for $\theta=\pi$.
Then, it turns out that for $\theta=\pi$, the pair of brains cannot move independently.

The Wilson loop also affects the mass of the states described by
\begin{align}
(\sigma_{3}\pm\i\sigma_{1})\Vgauge\e^{\i\frac{m}{R}X^{9}}.
\end{align}
The mass formula is the same as Eq.~\eqref{eq:mass_wboson}.
Since $\sigma_{1}$ and $\sigma_{3}$ are symmetric, the states with $m=1$ and $m=-1$ survive the orientifold projection.
In the case of $\theta=\pi$, two independent massless gauge fields appear.
Combining them to $\sigma_{2}\Vgauge$, the gauge symmetry enhances to $\SU(2)$.
This brane is the same as one obtained from the world-sheet with the symplectic Chan-Paton factor.

The above argument suggests the following interpretation of the group $\KR(S^{4}\times S^{1},S^{1})\simeq\BZ$.
The single wrapped $\Dp{4}$-brane obtained from the world-sheet after removing the anomaly gives a representative of the generator of the group $\KR(S^{4}\times S^{1},S^{1})$.
The pair of this D-brane with the orthogonal Chan-Paton factor corresponds to twice of the generator.
The wrapped $\Dp{4}$-brane formulated by the boundary of the world-sheet with the symplectic Chan-Paton factor gives rise to the representative of the same class of $\KR(S^{4}\times S^{1},S^{1})$ as this pair of branes.

\paragraph{D3-branes:}
The three Majorana fermions $\rho_{1}$, $\rho_{2}$, and $\rho_{3}$ that satisfy $\time\rho_{i}\time^{-1}=-\rho_{i}$ are added to formulate the wrapped $\Dp{3}$-brane.
Since there are tachyon vertex operators $\rho_{i}\Vtach$, the wrapped $\Dp{3}$-brane in this description is unstable.
After shifting the anomaly by $4$ units, the single Majorana fermion $\lambda$, that is even under $\time$, is relevant for coupling to the boundary.
Then, the situation is completely the same as the wrapped $\Dp{7}$-brane.
The state corresponding to the vertex operator~\eqref{eq:tach_D7} renders the brane unstable for $R<\sqrt{2\slope}$.
For $R>\sqrt{2\slope}$, this brane gives the representative of the class generating the group $\KR(S^{5}\times S^{1},S^{1})\simeq\BZ_{2}$.

\paragraph{D2-branes:}
We introduce two Majorana fermions $\rho_{1}$ and $\rho_{2}$ satisfying $\time\rho_{i}\time^{-1}=-\rho_{i}$ for the wrapped $\Dp{2}$-brane.
This brane is unstable because of the tachyon vertex operators $\rho_{i}\Vtach$.
In the description obtained by the change of the anomaly by the anomalous winding shift, we add two Majorana fermions $\lambda_{1}$ and $\lambda_{2}$ that are even under $\time$.
Similar to the argument for the wrapped $\Dp{6}$-brane, we can construct the tachyon vertex operator for any radius $R$ by appropriately turning on the Wilson loop.
These observations match the fact $\KR(S^{6}\times S^{1},S^{1})\simeq 0$.

\paragraph{D1-branes:}
To define the wrapped $\Dp{1}$-brane, we cancel the anomaly from the world-sheet fermions with a single $\time$-odd Majorana fermion on the boundary.
This Majorana fermion can be used to construct the tachyon vertex operator $\rho\Vtach$.
Then, this brane corresponds to the trivial class of the relative KR-theory group.
Even after moving on to the other description where three $\time$-even Majorana fermions $\lambda_{1}$, $\lambda_{2}$, and $\lambda_{3}$ are introduced, we can construct the tachyon vertex operator $\lambda_{1}\lambda_{2}\lambda_{3}\Vtach$, and the brane in this description also yields the trivial class.

\section{Boundary states}

\subsection{Boundary state formalism}\label{sub:boundary_state}
In this subsection, we confirm our convention concerning the boundary state formalism~\cite{Polchinski:1987tu,Callan:1988wz}.
See~\cite{Gaberdiel:2000jr} and Appendix~B of Ref.~\cite{Kaidi:2019tyf} for the review.

\subsubsection{Boundary state}
We begin with the boundary states $\ket{\Bp{p},\eta,\N}$ and $\ket{\Bp{p},\eta,\D}$ that constitute the $\Dp{p}$ brane state.
In the open string description, $\eta=\pm$ specifies the boundary conditions imposed on the world-sheet fermions.
From the ten-dimensional perspective, the $\Dp{p}$-branes in \Ndim{9} arise from the open string with both Neumann and Dirichlet boundary conditions for the compact direction $X^{9}$.
These two contributions are distinguished by the third component of the boundary states: the state $\ket{\Bp{p},\eta,\N}$ corresponds to the Neumann boundary condition for $X^{9}$, while $\ket{\Bp{p},\eta,\D}$ corresponds to the Dirichlet boundary condition for $X^{9}$.
The boundary states are characterized by several conditions involving the oscillator modes.
The conditions for the non-zero modes are given by
\begin{align}\begin{aligned}
	&\N:\quad(\alpha^{\mu}_{n}+\antialpha^{\mu}_{-n})\ket{\Bp{p},\eta,\N/\D}=(\psi^{\mu}_{r}+\i\eta\antipsi^{\mu}_{-r})\ket{\Bp{p},\eta,\N/\D}=0,\\
	&\D:\quad(\alpha^{l}_{n}-\antialpha^{l}_{-n})\ket{\Bp{p},\eta,\N/\D}=(\psi^{l}_{r}-\i\eta\antipsi^{l}_{-r})\ket{\Bp{p},\eta,\N/\D}=0.
\end{aligned}\end{align}
The first line represents the conditions for the modes in the direction with the Neumann boundary condition, and the second line represents those for the Dirichlet boundary condition.
For simplicity, we focus on the $\Dp{p}$-brane located at $X^{p+1}=\cdots=X^{8}=0$.
Then, the momentum $p^{\mu}$ along the Neumann direction and the center of mass $x^{l}$ along the Dirichlet direction yield the conditions:
\begin{align}\begin{aligned}\label{eq:bs_zero}
	&\N:\quad p^{\mu}\ket{\Bp{p},\eta,\N/\D}=0,\\
	&\D:\quad x^{l}\ket{\Bp{p},\eta,\N/\D}=0.
\end{aligned}\end{align}
In addition to these conditions from the zero-modes along the non-compact direction, the state $\ket{\Bp{p},\eta,\N}$ does not carry the KK momentum.
On the other hand, the state $\ket{\Bp{p},\eta,\D}$ consists of ones with zero winding number.

We choose the state,
\begin{align}
	\ket{\Bp{p},\eta,\N/\D}&=\int\frac{\d^{8-p}k}{(2\pi)^{8-p}}\exp\left[
		\sum_{n=1}^{\infty}\left(-\frac{1}{n}\sum_{\mu:\text{Neumann}}\alpha^{\mu}_{-n}\antialpha^{\mu}_{-n}+\frac{1}{n}\sum_{l:\text{Dirichlet}}\alpha^{l}_{-n}\antialpha^{l}_{-n}\right)
	\right.\notag\\
	&\hspace{0.6 in}\left.
		+\i\eta\sum_{r>0}^{\infty}\left(-\sum_{\mu:\text{Neumann}}\psi^{\mu}_{-r}\antipsi^{\mu}_{-r}+\sum_{l:\text{Dirichlet}}\psi^{l}_{-r}\antipsi^{l}_{-r}\right)
	\right]
	\ket{\Bp{p},\eta,\N/\D}^{(0)},
\end{align}
as the solution of all the conditions, where $\ket{\Bp{p},\eta,\N/\D}^{(0)}$ is the ground state.
The ground state includes the momentum eigenstate along the non-compact direction.
To satisfy the condition~\eqref{eq:bs_zero}, the momentum in the Neumann direction is set to zero, while that of the Dirichlet direction is integrated.
For convenience in notation, the dependence of the momentum along the non-compact directions is suppressed from the ground state $\ket{\Bp{p},\eta,\N/\D}^{(0)}$.
Since the ground states depend on the sector, we denote the sector as the subscript.
For the NSNS sector, the state $\ket{\Bp{p},\eta,\N/\D}^{(0)}_{\NS\NS}$ consists of the NSNS ground state $\ket{0}_{\NS\NS}$.
Following~\cite{Gaberdiel:2000jr}, the ground state for the RR sector is defined to satisfy the conditions,
\begin{align}\begin{aligned}
	&\N:\quad(\psi^{\mu}_{0}+\i\eta\antipsi^{\mu}_{0})\ket{\Bp{p},\eta}^{(0)}_{\R\R}=0,\\
	&\D:\quad(\psi^{l}_{0}-\i\eta\antipsi^{l}_{0})\ket{\Bp{p},\eta}^{(0)}_{\R\R}=0.
\end{aligned}\end{align}
The states $\ket{\Bp{p},\eta,\N}^{(0)}_{\NS\NS}$ and $\ket{\Bp{p},\eta,\N}^{(0)}_{\R\R}$ of the Neumann boundary condition for $X^{9}$ include
\begin{align}
	\frac{1}{2\pi R}\sum_{n=-\infty}^{\infty}\Ket{p_{9}=0,w_{9}=\frac{nR}{\slope}},
\end{align}
where $\ket{p_{9},w_{9}}$ is the momentum eigenstate along $X^{9}$ with the KK momentum $p_{9}$ and the winding number $w_{9}$, normalized as
\begin{align}
	\Braket{p_{9}=\frac{m}{R},w_{9}=\frac{nR}{\slope}|p_{9}=\frac{m'}{R},w_{9}=\frac{n'R}{\slope}}=2\pi R\,\delta_{m,m'}\delta_{n,n'}.
\end{align}
The states $\ket{\Bp{p},\eta,\D}^{(0)}_{\NS\NS}$ and $\ket{\Bp{p},\eta,\D}^{(0)}_{\R\R}$ of the Dirichlet boundary condition involve
\begin{align}
	\frac{1}{2\pi R}\sum_{m=-\infty}^{\infty}\e^{\i\frac{m}{R}x^{9}}\Ket{p_{9}=\frac{m}{R},w_{9}=0},
\end{align}
where $x^9$
represents the position of the D-brane along $X^{9}$.
We sometimes display the dependence on the position explicitly as $\ket{\Bp{p},\eta,\D,x^{9}}_{\NS\NS}$ and $\ket{\Bp{p},\eta,\D,x^{9}}_{\R\R}$.
In summary, the ground states are given by
\begin{align}\begin{aligned}
	\ket{\Bp{p},\eta,\N}^{(0)}_{\NS\NS}&=\ket{0}_{\NS\NS}\otimes\ket{k^{\mu}=0,k^{l}}\otimes\frac{1}{2\pi R}\sum_{n=-\infty}^{\infty}\Ket{p_{9}=0,w_{9}=\frac{nR}{\slope}},\\
	\ket{\Bp{p},\eta,\D,x^{9}}^{(0)}_{\NS\NS}&=\ket{0}_{\NS\NS}\otimes\ket{k^{\mu}=0,k^{l}}\otimes\frac{1}{2\pi R}\sum_{m=-\infty}^{\infty}\e^{\i\frac{m}{R}x^{9}}\Ket{p_{9}=\frac{m}{R},w_{9}=0},\\
	\ket{\Bp{p},\eta,\N}^{(0)}_{\R\R}&=2^{2}\ket{\Bp{(p+1)},\eta}_{\R\R}\otimes\ket{k^{\mu}=0,k^{l}}\otimes\frac{1}{2\pi R}\sum_{n=-\infty}^{\infty}\Ket{p_{9}=0,w_{9}=\frac{nR}{\slope}},\\
	\ket{\Bp{p},\eta,\D,x^{9}}^{(0)}_{\R\R}&=2^{2}\ket{\Bp{p},\eta}_{\R\R}\otimes\ket{k^{\mu}=0,k^{l}}\otimes\frac{1}{2\pi R}\sum_{m=-\infty}^{\infty}\e^{\i\frac{m}{R}x^{9}}\Ket{p_{9}=\frac{m}{R},w_{9}=0}.
\end{aligned}\end{align}
For later convenience, we include the additional factor $2^{2}$ in the definition of the ground states of the RR sector.

We introduce the normalization factors $\Dnorm{p}{\N}$ and $\Dnorm{p}{\D}$.
These normalization factors are determined by the relations:
\begin{align}
	\frac{1}{\Dnorm{p}{\N/\D}^{2}}\int_{0}^{\infty}\,\d s\,{}_{\NS\NS}\bra{\Bp{p},\eta,\N/\D}\e^{-2\pi sH_{\mathrm{c}}}\ket{\Bp{p},\eta,\N/\D}_{\NS\NS}&=\int_{0}^{\infty}\,\frac{\d t}{2t}\,\Tr_{\NS}\e^{-2\pi tH_{\mathrm{o}}},\notag\\
	\frac{1}{\Dnorm{p}{\N/\D}^{2}}\int_{0}^{\infty}\,\d s\,{}_{\NS\NS}\bra{\Bp{p},\eta,\N/\D}\e^{-2\pi sH_{\mathrm{c}}}\ket{\Bp{p},-\eta,\N/\D}_{\NS\NS}&=\int_{0}^{\infty}\,\frac{\d t}{2t}\,\Tr_{\R}\e^{-2\pi tH_{\mathrm{o}}},\\
	\frac{1}{\Dnorm{p}{\N/\D}^{2}}\int_{0}^{\infty}\,\d s\,{}_{\R\R}\bra{\Bp{p},\eta,\N/\D}\e^{-2\pi sH_{\mathrm{c}}}\ket{\Bp{p},\eta,\N/\D}_{\R\R}&=\int_{0}^{\infty}\,\frac{\d t}{2t}\,\Tr_{\NS}(-1)^{\fermion}\e^{-2\pi tH_{\mathrm{o}}},\notag
\end{align}
where $H_{\mathrm{c}}$ and $H_{\mathrm{o}}$ are Hamiltonians of the closed string and the open string, respectively.
The explicit expressions are given by
\begin{align}\begin{aligned}
	\Dnorm{p}{\N}&=2^{\frac{5}{2}}(4\pi^{2}\slope)^{\frac{2p-7}{4}}\left(\frac{\slope}{R^{2}}\right)^{\frac{1}{4}}(2\pi R)^{-\frac{1}{2}},\\
	\Dnorm{p}{\D}&=2^{\frac{5}{2}}(4\pi^{2}\slope)^{\frac{2p-7}{4}}\left(\frac{R^{2}}{\slope}\right)^{\frac{1}{4}}(2\pi R)^{-\frac{1}{2}}.
\end{aligned}\end{align}

The left-moving fermion number $(-1)^{\fermion_{\L}}$ and the right-moving fermion number $(-1)^{\fermion_{\R}}$ act on the boundary states of the NSNS sector as
\begin{align}\begin{aligned}
	(-1)^{\fermion_{\L}}\ket{\Bp{p},\eta,\N/\D}_{\NS\NS}&=-\ket{\Bp{p},-\eta,\N/\D}_{\NS\NS},\\
	(-1)^{\fermion_{\R}}\ket{\Bp{p},\eta,\N/\D}_{\NS\NS}&=-\ket{\Bp{p},-\eta,\N/\D}_{\NS\NS},
\end{aligned}\end{align}
because the NSNS ground state has the eigenvalue $-1$ for these operators.
On the other hand, the actions of two fermion number operators act on the boundary states of the RR sector are given by
\begin{align}
	(-1)^{\fermion_{\L}}\ket{\Bp{p},\eta,\N}_{\R\R}&=(-1)^{6-p}\ket{\Bp{p},-\eta,\N}_{\R\R},\\
	(-1)^{\fermion_{\L}}\ket{\Bp{p},\eta,\D}_{\R\R}&=(-1)^{7-p}\ket{\Bp{p},-\eta,\D}_{\R\R},\\
	(-1)^{\fermion_{\R}}\ket{\Bp{p},\eta,\N/\D}_{\R\R}&=\ket{\Bp{p},-\eta,\N/\D}_{\R\R}.
\end{align}
We define the $\Dp{p}$-brane state as
\begin{align}\begin{aligned}\label{eq:brane_state}
	\ket{\Dp{p},\N/\D}=\frac{1}{2\sqrt{2}\Dnorm{p}{\N/\D}}\left[
		(\ket{\Bp{p},+,\N/\D}_{\NS\NS}-\ket{\Bp{p},-,\N/\D}_{\NS\NS})
	\right.\\
	\left.
		+(\ket{\Bp{p},+,\N/\D}_{\R\R}+\ket{\Bp{p},-,\N/\D}_{\R\R})
	\right].
\end{aligned}\end{align}
For the Dirichlet boundary condition, there is the additional parameter $x^{9}$ that describes the position in the $X^{9}$ direction.
Note that the state $\ket{\Dp{p},\N}$ is invariant under the GSO projection for even $p$, while $\ket{\Dp{p},\D}$ is invariant for odd $p$.

\subsubsection{Crosscap state}
To discuss the D-brane in the orientifold theory, we need the crosscap states $\ket{\C,\eta}$.
These states are defined as
\begin{align}
	\ket{\C,\eta}_{\NS\NS}=\i\exp\left[
		-\sum_{\mu=2}^{9}\left(\sum_{n=1}^{\infty}\frac{1}{n}(-1)^{n}\alpha^{\mu}_{-n}\antialpha^{\mu}_{-n}
		+\i\eta\sum_{r>0}^{\infty}e^{\i\pi r}\psi^{\mu}_{-r}\antipsi^{\mu}_{-r}\right)
	\right]
	\ket{\C,\eta}^{(0)}_{\NS\NS},
\end{align}
where the ground state is given by
\begin{align}\begin{aligned}
	\ket{\C,\eta}^{(0)}_{\NS\NS}&=\ket{0}_{\NS\NS}\otimes\ket{k^{\mu}=0}\otimes\sum_{n=-\infty}^{\infty}\Ket{p_{9}=0,w_{9}=\frac{(2n+1)R}{\slope}}.\\
\end{aligned}\end{align}
Since the crosscap states in the RR sector do not play any role in the following argument, we do not show these states here.

For the normalization factor $\Cnorm=2^{-5}\Dnorm{8}{\N}$, the relations
\begin{align}
	\frac{1}{\Cnorm^{2}}\int_{0}^{\infty}\,\d s\,{}_{\NS\NS}\bra{\C,\eta}\e^{-2\pi sH_{\mathrm{c}}}\ket{\C,\eta}_{\NS\NS}&=\int_{0}^{\infty}\,\frac{\d t}{2t}\,\Tr_{\NS\NS}S\Omega\e^{-2\pi tH_{\mathrm{c}}},\\
	\frac{1}{\Cnorm^{2}}\int_{0}^{\infty}\,\d s\,{}_{\NS\NS}\bra{\C,\eta}\e^{-2\pi sH_{\mathrm{c}}}\ket{\C,-\eta}_{\NS\NS}&=\int_{0}^{\infty}\,\frac{\d t}{2t}\,\Tr_{\R\R}S\Omega\e^{-2\pi tH_{\mathrm{c}}},
\end{align}
are satisfied.
The total crosscap state is given by the linear combination as
\begin{align}
	\ket{\Op{}}=\frac{1}{2\sqrt{2}\Cnorm}
		(\ket{\C,+}_{\NS\NS}-\ket{\C,-}_{\NS\NS})+(\text{crosscap states in the RR sector}),
\end{align}
so that the overlap
\begin{align}
	\int_{0}^{\infty}\,\d s\,\bra{\Op{}}\e^{-2\pi sH_{\mathrm{c}}}\ket{\Op{}}
\end{align}
provides the Klein bottle amplitude.

\subsection{Unwrapped brane}\label{sub:state_unwrapped}
Since the orientifold action of the $\Itilde$ background involves the half-shift $S$, the mirror image is needed to make the D-brane state invariant.
For $p=1$ and $p=5$, the mirror image of the $\Dp{p}$-branes are themselves.
Then, the states
\begin{align}
	\ket{\Dp{p},\D,x^{9}=0}+\ket{\Dp{p},\D,x^{9}=\pi R}
\end{align}
become invariant under both the GSO projection and the orientifold, and generate the groups $\KR(S^{3}\times S^{1},S^{1})\simeq\BZ$ and $\KR(S^{7}\times S^{1},S^{1})\simeq\BZ$.

For $p=3$ and $p=7$, the mirror image of the $\Dp{p}$-brane are antibranes.
The antibrane states are obtained by flipping the sign of the RR states in Eq.~\eqref{eq:brane_state}, that is
\begin{align}\begin{aligned}
	\ket{\overline{\Dp{p}},\D}=\frac{1}{2\sqrt{2}\Dnorm{p}{\N/\D}}\left[
		(\ket{\Bp{p},+,\N/\D}_{\NS\NS}-\ket{\Bp{p},-,\N/\D}_{\NS\NS})
	\right.\\
	\left.
		-(\ket{\Bp{p},+,\N/\D}_{\R\R}+\ket{\Bp{p},-,\N/\D}_{\R\R})
	\right].
\end{aligned}\end{align}
The invariant states are constructed by
\begin{align}
	\ket{\Dp{\widetilde{p}},\D}:=\ket{\Dp{p},\D,x^{9}=0}+\ket{\overline{\Dp{p}},\D,x^{9}=\pi R}.
\end{align}
To discuss the stability of these branes, we calculate the cylinder amplitude,
\begin{align}
	\mathcal{A}_{C_{2}}&=\int_{0}^{\infty}\,\d s\,\bra{\Dp{\widetilde{p}},\D}\e^{-2\pi sH_{\mathrm{c}}}\ket{\Dp{\widetilde{p}},\D}\notag\\
	&=\frac{1}{2}V_{p+1}(4\pi^{2}\slope)^{-\frac{p+1}{2}}\int_{0}^{\infty}\,\frac{\d t}{(2t)^{\frac{p+3}{2}}}\,\eta(\i t)^{-12}\notag\\
	&\hspace{0.6 in}\times\left[
		\vartheta_{00}\left(\frac{2\i R^{2}t}{\slope}\right)\left(\vartheta_{00}(\i t)^{4}-\vartheta_{01}(\i t)^{4}\right)+\vartheta_{10}\left(\frac{2\i R^{2}t}{\slope}\right)\left(\vartheta_{00}(\i t)^{4}+\vartheta_{01}(\i t)^{4}\right)
	\right.\notag\\
	&\hspace{3.6 in}\left.
		-\vartheta_{00}\left(\frac{2\i R^{2}t}{\slope}\right)\vartheta_{10}(\i t)^{4}
	\right],
\end{align}
where $V_{p+1}$ is the volume of the brane in terms of \Ndim{9}.
The convention of the theta functions are summarized in Eq.~\eqref{eq:theta}.
The contribution from the lightest modes of the open string can be read off as
\begin{align}
	\mathcal{A}_{C_{2}}|_{\text{lightest mode}}
	=2V_{p+1}(4\pi^{2}\slope)^{-\frac{p+1}{2}}\int_{0}^{\infty}\,\frac{\d t}{(2t)^{\frac{p+3}{2}}}\,\e^{\pi\left(1-\frac{R^{2}}{2\slope}\right)t}.
\end{align}
This result implies that for $R<\sqrt{2\slope}$, the brane is unstable.
On the other hand, for $R>\sqrt{2\slope}$, the tachyonic modes are absent, and the branes are the representatives of the generators of the groups $\KR(S^{1}\times S^{1},S^{1})\simeq\BZ_{2}$ and $\KR(S^{5}\times S^{1},S^{1})\simeq\BZ_{2}$

\subsection{Wrapped brane}\label{sub:state_wrapped}
For $p=4$ and $p=8$, the D-brane state~\eqref{eq:brane_state} is invariant not only under the GSO projection but also the orientifold projection.
These branes have the integer valued charges of the massless RR fields.

To construct the non-BPS $\Dp{3}$-branes and $\Dp{7}$-brane observed in subsection~\ref{sub:wrapped}, we consider the brane-antibrane system,
\begin{align}\label{eq:nonbps}
	\ket{\Dp{\widetilde{p}},\N}:=\lambda_{p}\left(\ket{\Dp{p},\N}+\ket{\overline{\Dp{p}},\N}\right)=
	\frac{\lambda_{p}}{\sqrt{2}\Dnorm{p}{\N/\D}}\left(\ket{\Bp{p},+,\N}_{\NS\NS}-\ket{\Bp{p},-,\N}_{\NS\NS}\right),
\end{align}
where the factor $\lambda_{p}>0$ is determined below.
Since the states in the RR sector do not appear in this expression, this state~\eqref{eq:nonbps} is invariant under the orientifold projection for any $p$.
However, it follows from the absence of the RR sector that the GSO projection is not taken in the open string perspective.
As the result, the cylinder amplitude has the contribution from the tachyonic modes.
The total amplitude is given by
\begin{align}
	\mathcal{A}_{C_{2}}&=\int_{0}^{\infty}\,\d s\,\bra{\Dp{\widetilde{p}},\N}\e^{-2\pi sH_{\mathrm{c}}}\ket{\Dp{\widetilde{p}},\N}\notag\\
	&=\lambda_{p}^{2}V_{p+1}(4\pi^{2}\slope)^{-\frac{p+1}{2}}\int_{0}^{\infty}\,\frac{\d t}{(2t)^{\frac{p+3}{2}}}\,\eta(\i t)^{-12}\vartheta_{00}\left(\frac{2\i \slope t}{R^{2}}\right)\left(\vartheta_{00}(\i t)^{4}-\vartheta_{10}(\i t)^{4}\right)\notag\\
	&=\lambda_{p}^{2}V_{p+1}(4\pi^{2}\slope)^{-\frac{p+1}{2}}\int_{0}^{\infty}\,\frac{\d t}{(2t)^{\frac{p+3}{2}}}\,\vartheta_{00}\left(\frac{2\i \slope t}{R^{2}}\right)\e^{\pi t}+\cdots,
\end{align}
where we extract the part that may diverge in the last line.
Note that in the $\Itilde$ background we need to take into account the M\"{o}bius amplitude, and it may be possible that the tachyonic modes in the cylinder amplitude is canceled by that in the M\"{o}bius amplitude.
The M\"{o}bius amplitude can be calculated as
\begin{align}
	\mathcal{A}_{M_{2}}
	&=\int_{0}^{\infty}\,\d s\,\left(\bra{\Op{}}\e^{-2\pi sH_{\mathrm{c}}}\ket{\Dp{\widetilde{p}},\N}+\bra{\Dp{\widetilde{p}},\N}\e^{-2\pi sH_{\mathrm{c}}}\ket{\Op{}}\right)\notag\\
	&=-\frac{\lambda_{p}}{2}V_{p+1}(4\pi^{2}\slope)^{-\frac{p+1}{2}}\int_{0}^{\infty}\,\frac{\d t}{(2t)^{\frac{p+3}{2}}}\,\vartheta_{01}\left(\frac{2\i \slope t}{R^{2}}\right)\notag\\
	&\hspace{0.4 in}\times\left[
		\e^{-\i\frac{\pi}{4}p}\left(f_{1}^{-p}f_{3}^{8}f_{4}^{2(8-p)}\right)\left(\i t+\frac{1}{2}\right)-\e^{\i\frac{\pi}{4}p}\left(f_{1}^{-p}f_{3}^{2(8-p)}f_{4}^{8}\right)\left(\i t+\frac{1}{2}\right)
	\right]\notag\\
	&=\lambda_{p}V_{p+1}(4\pi^{2}\slope)^{-\frac{p+1}{2}}\sin\left(\frac{\pi}{4}p\right)\int_{0}^{\infty}\,\frac{\d t}{(2t)^{\frac{p+3}{2}}}\,\vartheta_{01}\left(\frac{2\i \slope t}{R^{2}}\right)\e^{\pi t}+\cdots.
\end{align}
The definitions of the functions $f_{1}(\tau)$, $f_{3}(\tau)$, and $f_{4}(\tau)$ are listed in Eq.~\eqref{eq:f_func}.
Then, the total contribution is
\begin{align}
	\mathcal{A}_{C_{2}}+\mathcal{A}_{M_{2}}
	&=\lambda_{p}V_{p+1}(4\pi^{2}\slope)^{-\frac{p+1}{2}}\left(\lambda_{p}+\sin\left(\frac{\pi}{4}p\right)\right)\int_{0}^{\infty}\,\frac{\d t}{(2t)^{\frac{p+3}{2}}}\,\e^{\pi t}\sum_{n=-\infty}^{\infty}\e^{-\frac{4\pi\slope t}{R^{2}}\frac{(2n)^{2}}{2}}\notag\\
	&\quad+\lambda_{p}V_{p+1}(4\pi^{2}\slope)^{-\frac{p+1}{2}}\left(\lambda_{p}-\sin\left(\frac{\pi}{4}p\right)\right)\int_{0}^{\infty}\,\frac{\d t}{(2t)^{\frac{p+3}{2}}}\,\e^{\pi t}\sum_{n=-\infty}^{\infty}\e^{-\frac{4\pi\slope t}{R^{2}}\frac{(2n+1)^{2}}{2}}\notag\\
	&\quad+\cdots.
\end{align}
The first line can be interpreted as the contribution from the open string with the even KK momentum, while the second line corresponds to that from the open string with the odd KK momentum.
For $p=5,6,7$, the contribution from the even KK momentum can vanish by setting $\lambda_{p}=-\sin\frac{\pi}{4}p>0$.
The open string with the unit KK momentum that is the lightest mode becomes massive for $R<\sqrt{2\slope}$.
This wrapped $\Dp{7}$-brane is one constructed in section~\ref{sub:wrapped}.
In the boundary state formalism, it seems to be possible to construct $\Dp{5}$-brane and $\Dp{6}$-brane.
However, as discussed in section~\ref{sub:wrapped}, the wrapped $\Dp{6}$-brane becomes unstable by turning on the Wilson loop.
For the wrapped $\Dp{5}$-brane, we can write down the tachyon vertex operator.
The same phenomena also occur in the type~I theory~\cite{Frau:1999qs} and the type~0 orientifold theory~\cite{Kaidi:2019tyf} in \Ndim{10}.
In these theories, non-BPS $\Dp{6}$-brane seems to be allowed in terms of the boundary state formalism.
However, it follows from the argument in Ref.~\cite{Gimon:1996rq} that the tachyon in the system of $\Dp{6}$-brane and $\overline{\Dp{6}}$-brane actually cannot be removed by the orientifold projection.

As discussed in section~\ref{sub:mixed_anomaly}, we can freely add the topological term~\eqref{eq:inv_phase} on the world-sheet.
In the presence of this term, the sign of the M\"{o}bius amplitude is flipped.
As the result, the possible divergent part is changed as
\begin{align}
	\mathcal{A}_{C_{2}}+\mathcal{A}_{M_{2}}
	&=\lambda_{p}V_{p+1}(4\pi^{2}\slope)^{-\frac{p+1}{2}}\left(\lambda_{p}-\sin\left(\frac{\pi}{4}p\right)\right)\int_{0}^{\infty}\,\frac{\d t}{(2t)^{\frac{p+3}{2}}}\,\e^{\pi t}\sum_{n=-\infty}^{\infty}\e^{-\frac{4\pi\slope t}{R^{2}}\frac{(2n)^{2}}{2}}\notag\\
	&\quad+\lambda_{p}V_{p+1}(4\pi^{2}\slope)^{-\frac{p+1}{2}}\left(\lambda_{p}+\sin\left(\frac{\pi}{4}p\right)\right)\int_{0}^{\infty}\,\frac{\d t}{(2t)^{\frac{p+3}{2}}}\,\e^{\pi t}\sum_{n=-\infty}^{\infty}\e^{-\frac{4\pi\slope t}{R^{2}}\frac{(2n+1)^{2}}{2}}\notag\\
	&\quad+\cdots.
\end{align}
In this case, the divergence from the even KK momentum can be canceled for $p=1,2,3$.
Similar to the argument in the previous paragraph, only the $\Dp{3}$-brane is actually stable for $R<\sqrt{2\slope}$.

\subsection*{Acknowledgement}
We are grateful to Shoto Aoki and Piljin Yi for the valuable discussions.
We also thank the Yukawa Institute for Theoretical Physics at Kyoto University. 
Discussions during YITP workshop ``Strings and Fields 2024'' (YITP-W-24-08) and YITP-RIKEN iTHEMS conference ``Generalized symmetries in QFT 2024'' (YITP-W-24-15) are useful to complete this work.
H.W. acknowledges the hospitality at Korea Institute for Advanced Study during his stay.
The work of H.W. was supported in part by JSPS KAKENHI Grant-in-Aid for JSPS fellows Grant Number 24KJ1603.  The work of S.Y. was supported in part by JSPS KAKENHI Grant Number 21K03574.

\appendix
\section{Theta functions}\label{sec:theta}
The theta functions used in this paper are defined as
\begin{align}\begin{aligned}\label{eq:theta}
    \vartheta_{00}(\nu,\tau)&=\vartheta_{3}(\nu\,|\,\tau)=\sum_{n=-\infty}^{\infty}q^{n^{2}/2}z^{n},\\
    \vartheta_{01}(\nu,\tau)&=\vartheta_{4}(\nu\,|\,\tau)=\sum_{n=-\infty}^{\infty}(-1)^{n}q^{n^{2}/2}z^{n},\\
    \vartheta_{10}(\nu,\tau)&=\vartheta_{2}(\nu\,|\,\tau)=\sum_{n=-\infty}^{\infty}q^{(n-1/2)^{2}/2}z^{n-1/2},\\
    \vartheta_{11}(\nu,\tau)&=-\vartheta_{1}(\nu\,|\,\tau)=-\i\sum_{n=-\infty}^{\infty}(-1)^{n}q^{(n-1/2)^{2}/2}z^{n-1/2},
\end{aligned}\end{align}
where
\begin{align}
    q=\exp(2\pi\i\tau),\quad z=\exp(2\pi\i\nu).
\end{align}
We introduce the notation $\vartheta_{\alpha\beta}(\tau)=\vartheta_{\alpha\beta}(\nu=0,\tau)$.
The Dedekind eta function is
\begin{align}
    \eta(\tau)=q^{1/24}\prod_{m=1}^{\infty}(1-q^{m}).
\end{align}
We also utilize the following combinations of the theta functions and the Dedekind eta function:
\begin{align}\begin{aligned}\label{eq:f_func}
    f_{1}(\tau)&=\eta(\tau)=q^{1/24}\prod_{m=1}^{\infty}(1-q^{m}),\\
    f_{2}(\tau)&=\sqrt{\frac{\vartheta_{2}(0,\tau)}{\eta(\tau)}}=\sqrt{2}q^{1/24}\prod_{m=1}^{\infty}(1+q^{m}),\\
    f_{3}(\tau)&=\sqrt{\frac{\vartheta_{3}(0,\tau)}{\eta(\tau)}}=q^{-1/48}\prod_{m=1}^{\infty}(1+q^{m-1/2}),\\
    f_{4}(\tau)&=\sqrt{\frac{\vartheta_{4}(0,\tau)}{\eta(\tau)}}=q^{-1/48}\prod_{m=1}^{\infty}(1-q^{m-1/2}).
\end{aligned}\end{align}

\section{Mode expansion}
This section is devoted to confirm the notation about the world-sheet theory.
We mainly follow the convention used in Ref.~\cite{Kaidi:2019tyf}.
In the conformal gauge, the fundamental fields of the world-sheet theory are bosons $X^{\mu}$, and two component Majorana fermions.
Except for section~\ref{sec:anomaly}, we denote the left-moving and right-moving component of the fermions as $\psi^{\mu}$ and $\antipsi^{\mu}$, respectively.
In the $\Itilde$ background, the $X^{9}$ direction is compactified on a circle $S^{1}$ with radius $R$.
We adopt the light-cone gauge, and choose $X^{0}$ and $X^{1}$ as the light-cone coordinates.
\subsection{Closed string}
Let us parametrize the cylinder by $t$ and $\sigma$.
We regard $t$ as the time direction and $\sigma\in[0,2\pi)$ as the spatial direction of the world-sheet.
The action of the world-sheet theory is written as
\begin{align}
    S=\frac{1}{4\pi}\int\,\d t\,\d\sigma\left[-\frac{1}{\slope}\left(-\del_{t}X^{\mu}\del_{t}X_{\mu}+\del_{\sigma}X^{\mu}\del_{\sigma}X_{\mu}\right)+\i\psi^{\mu}(\del_{t}+\del_{\sigma})\psi_{\mu}+\i\antipsi^{\mu}(\del_{t}-\del_{\sigma})\antipsi_{\mu}\right].
\end{align}
The oscillator expansions for the bosons are given by
\begin{align}
	X^{\mu}(t,\sigma)&=x^{\mu}+\frac{\slope}{2}p^{\mu}_{\L}(t-\sigma)+\frac{\slope}{2}p^{\mu}_{\R}(t+\sigma)+\i\sqrt{\frac{\slope}{2}}\sum_{m\neq 0}\frac{1}{m}\left(\alpha^{\mu}_{m}\e^{-\i m(t-\sigma)}+\antialpha^{\mu}_{m}\e^{-\i m(t+\sigma)}\right).
\end{align}
For non-compact direction $\mu=2,\dots,8$, the zero-modes $p^{\mu}=p^{\mu}_{\L}=p^{\mu}_{\R}$ are the momentum.
In $X^{9}$ direction, the zero-modes are related to the KK momentum $m\in\BZ$ and the winding number $n\in\BZ$,
\begin{align}
		p^{9}_{L}&=\frac{m}{R}-\frac{nR}{\slope},&
		p^{9}_{R}&=\frac{m}{R}+\frac{nR}{\slope}.
\end{align}
The mode expansions for the fermions are
\begin{align}
	\psi^{\mu}(t,\sigma)&=\sum_{r}\psi^{\mu}_{r}\e^{-\i r(t-\sigma)},&
	\antipsi^{\mu}(t,\sigma)&=\sum_{r}\antipsi^{\mu}_{r}\e^{-\i r(t+\sigma)}.
\end{align}
In terms of these oscillators, the Hamiltonian of the closed string is written as
\begin{align}
	H_{\mathrm{c}}=\frac{\slope}{2}p^{2}+\frac{\slope}{2}\left(\frac{m}{R}\right)^{2}+\frac{n^{2}R^{2}}{2\slope}+\sum_{n=1}^{\infty}\left(\alpha_{-n}\cdot\alpha_{n}+\antialpha_{-n}\cdot\antialpha_{n}\right)+\sum_{r>0}r\left(\psi_{-r}\psi_{r}+\antipsi_{-r}\antipsi_{r}\right)-a-\widetilde{a},
\end{align}
where $a$ and $\widetilde{a}$ are the vacuum energy of left-moving and right-moving parts, respectively.

The world-sheet parity acts on the fields as
\begin{align}\begin{aligned}
	\Omega X^{\mu}(t,\sigma)\Omega^{-1}&=X^{\mu}(t,2\pi-\sigma),\\
	\Omega\psi^{\mu}(t,\sigma)\Omega^{-1}=-\antipsi^{\mu}(t,2\pi-\sigma),&\quad\Omega\antipsi^{\mu}(t,\sigma)\Omega^{-1}=\psi^{\mu}(t,2\pi-\sigma).
\end{aligned}\end{align}

\subsection{Open string}
For the open string, we consider the world-sheet theory on a strip of a width $\pi$.
We take the coordinate so that $\sigma=0$ and $\sigma=\pi$ are the boundaries.
\paragraph{Neumann-Neumann boundary condition:}
For the world-sheet boson imposed the Neumann boundary condition for both ends, the mode expansion is given by
\begin{align}
	X^{\mu}(t,\sigma)=x^{\mu}+2\slope p^{\mu}t+\i\sqrt{\frac{\slope}{2}}\sum_{m\neq 0}\frac{1}{m}\alpha_{m}^{\mu}\left(e^{-\i m(t-\sigma)}+e^{-\i m(t+\sigma)}\right).
\end{align}
In the compact direction $X^{9}$, the zero-mode $p_{9}$ corresponds to the KK momentum.
For world-sheet fermions, we impose the boundary conditions\footnote{The boundary conditions~\eqref{eq:NN_bc} and~\eqref{eq:DD_bc} are chosen so that the two components of the supercurrent $G_{--}=\frac{1}{\sqrt{2\slope}}\psi(\del_{t}+\del_{\sigma})X$ and $G_{++}=\frac{1}{\sqrt{2\slope}}\antipsi(\del_{t}-\del_{\sigma})X$ match at $\sigma=\pi$.}
\begin{align}\label{eq:NN_bc}
	\psi^{\mu}(t,0)&=\eta\antipsi^{\mu}(t,0),&\psi^{\mu}(t,\pi)=\antipsi^{\mu}(t,\pi),
\end{align}
where $\eta=-1$ and $\eta=+1$ correspond to the NS sector and the R sector, respectively.
Since the left-moving fermions number $(-1)^{\fermion_{\L}}$ is gauged in the $\Itilde$ background, we can fix the boundary condition at $\sigma=\pi$ by the redefinition of $\psi^{\mu}$.
The oscillator expansions for the fermions are given by
\begin{align}
	\psi^{\mu}(t,\sigma)&=\sum_{r}\psi^{\mu}_{r}\e^{-\i r(t-\sigma)},&\antipsi^{\mu}(t,\sigma)&=\eta\sum_{r}\psi^{\mu}_{r}\e^{-\i r(t+\sigma)}.
\end{align}
The Hamiltonian of the open string is written as
\begin{align}\label{eq:open_Hamiltonian}
	H_{\mathrm{o}}=\slope p^{2}+\slope\left(\frac{m}{R}\right)^{2}+\sum_{n=1}^{\infty}\alpha_{-n}\cdot\alpha_{n}+\sum_{r>0}r\psi_{-r}\psi_{r}-a.
\end{align}

The actions of the world-sheet parity on the oscillators are deduced as
\begin{align}
	\Omega\alpha^{\mu}_{m}\Omega^{-1}&=(-1)^{m}\alpha^{\mu}_{m},&\Omega\psi^{\mu}_{r}\Omega^{-1}&=\e^{-\i\pi r}\psi^{\mu}_{r}.
\end{align}

\paragraph{Dirichlet-Dirichlet boundary condition:}
The mode expansion for the boson that satisfies the Dirichlet boundary condition on both edges is
\begin{align}
	X^{\mu}(t,\sigma)=\i\sqrt{\frac{\slope}{2}}\sum_{m\neq 0}\frac{1}{m}\alpha_{m}^{\mu}\left(-e^{-\i m(t-\sigma)}+e^{-\i m(t+\sigma)}\right).
\end{align}
For the compact direction $X^{9}$, the winding number $(2n R+\frac{\Delta x^{9}}{\pi})\sigma$ is added to the expansion, where $n\in\BZ$ and $\Delta x^{9}$ is the distance of two ends.
In this paper, the only situations that $\Delta x^{9}=0$ and $\Delta x^{9}=\pi R$ appear.
The boundary conditions for the world-sheet fermions are chosen as
\begin{align}\label{eq:DD_bc}
	\psi^{\mu}(t,0)&=-\eta\antipsi^{\mu}(t,0),&\psi^{\mu}(t,\pi)=-\antipsi^{\mu}(t,\pi).
\end{align}
The mode expansions are
\begin{align}
	\psi^{\mu}(t,\sigma)&=\sum_{r}\psi^{\mu}_{r}\e^{-\i r(t-\sigma)},&\antipsi^{\mu}(t,\sigma)&=-\eta\sum_{r}\psi^{\mu}_{r}\e^{-\i r(t+\sigma)}.
\end{align}
The Hamiltonian for this boundary condition is obtained by replacing the contribution from the zero-modes to $\frac{1}{4\pi^{2}\slope}(2\pi nR+\Delta x^{9})^{2}$ in Eq.~\eqref{eq:open_Hamiltonian}.

The actions of the world-sheet parity on the oscillators are given by
\begin{align}
	\Omega\alpha^{\mu}_{m}\Omega^{-1}&=-(-1)^{m}\alpha^{\mu}_{m},&\Omega\psi^{\mu}_{r}\Omega^{-1}&=-\e^{-\i\pi r}\psi^{\mu}_{r}.
\end{align}

\bibliographystyle{utphys}
\bibliography{Klein}

\end{document}